\documentclass[a4paper]{jpconf}
\usepackage[dvips]{graphicx,epsfig,color}
\usepackage{wrapfig,rotating}
\usepackage{amssymb,amsmath,array}

\newcommand{\la}{\langle}
\newcommand{\ra}{\rangle}
\newcommand{\ph}{{ \phi _h}}

\newcommand{\cs}{\la {\cos\ph} \ra}
\newcommand{\cst}{\la {\cos2\ph} \ra}

\newcommand{\csm}{\cos\ph}
\newcommand{\cstm}{\cos2\ph}

\newcommand{\hermes}{{\sc{Hermes}}}
\newcommand{\hera}{{\sc Hera}}
\newcommand{\desy}{{\sc desy}}
\newcommand{\compass}{{\sc Compass}}

\begin{document}
\title{Flavor-dependent azimuthal modulations in unpolarized SIDIS
  cross section at \hermes}

\author{Francesca Giordano$^1$ and Rebecca Lamb$^2$
\\On behalf of the  \hermes\ collaboration}

\address{$^1$ Deutsches Elektronen-Synchrotron, Notkestra\ss e 85,
22607 Hamburg, Germany}
\address{$^2$ University of Illinois,
Department of Physics, 1110 West Green Street, Urbana, IL 61801-3080, USA}

\ead{francesca.giordano@desy.de, rebecca.lamb@desy.de}

\begin{abstract}
The $\csm$ and $\cstm$ azimuthal modulations of the unpolarized hadron
Semi-Inclusive Deep Inelastic Scattering cross section are sensitive to the
quark intrinsic transverse momentum and transverse spin.
These modulations have been measured at \hermes\ in a fully differential way
by means of a 4-dimensional unfolding procedure to correct for instrumental
effects.
Results have been extracted for hydrogen and deuterium targets and
separately for positively and negatively charged pions and kaons, to access
flavor-dependent information about the nucleon internal transverse degrees of
freedom.
\end{abstract}

In lepton-nucleon deep-inelastic scattering (DIS), 
the structure of the nucleon is probed by the interaction of
a high energy lepton with a target nucleon,
via, at \hermes\ kinematics, the exchange of one virtual photon. If at least one of the 
produced hadrons is detected in coincidence with the scattered lepton, 
the reaction is called semi-inclusive deep-inelastic scattering (SIDIS):
\begin{equation}\label{eq:one}
l({\bf k})\,+\,N({\bf P})\, \rightarrow \, l'({\bf k}')\, + \,h({\bf P}_h)\,\,+\,X({\bf P}_X),
\end{equation}
where $l$ ($l'$) is the incident (scattered) lepton, 
$N$ is the target nucleon, $h$ is a detected hadron, $X$ is 
the target remnant and the quantities in parentheses in equation~(\ref{eq:one}) are 
the corresponding four-momenta.

\begin{figure}
\centering
\includegraphics[height=.2\textheight]{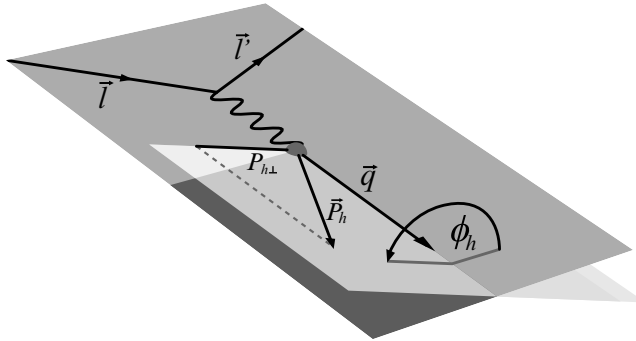}
\caption{Definition of the azimuthal angle $\phi_h$ between
scattering plane, spanned by the in- and out-going lepton three-momenta ($\vec{l}$, $\vec{l'}$),
and the hadron production plane, defined by the three-momenta of the
virtual photon ($\vec{q}$) and produced hadron ($\vec{P_h}$).}
\label{fig:evento}
\end{figure}

If unintegrated over the hadron momentum
component transverse to the virtual photon direction $P_{h\perp}$ (Fig.~\ref{fig:evento}),
the cross section can be written as~\cite{Bacchetta:2006tn}:
\begin{equation}
\centering 
\begin{split}
d\sigma\equiv\frac{d\sigma}{dx\,dy\,dz\,dP^2_{h\perp}\,d\phi_h}=
\frac{\alpha^2}{xyQ^2}(1+\frac{\gamma^2}{2x})
\{A(y)\,F_{UU,T}+B(y)\, F_{UU,L}+\\ 
C(y)\,\cos\phi _h F_{UU}^{\cos\phi _h}+B(y)\,\cos2\phi _h F_{UU}^{\cos2\phi _h}\},
\label{eq:noncol_csec}
\end{split}
\end{equation}
where $F_{UU}^{\cos\phi _h}$, $F_{UU}^{\cos2\phi _h}$, are azimuthally
dependent structure functions, and are related respectively to 
$\cos\phi _h$ and $\cos2\phi _h$ modulations, with
$\phi_h$ the azimuthal angle of the hadron production plane around the 
virtual-photon direction (Fig.~\ref{fig:evento}). 
In equation~\ref{eq:noncol_csec}, the subscripts $UU$ stand for unpolarized beam and target, 
$T$ ($L$) indicates the transverse (longitudinal) polarization of the virtual 
photon, $\alpha$ is the electromagnetic coupling constant, $\gamma=2Mx/Q$ 
with $M$ the target mass, $A(y) \sim (1-y+1/2y^2)$, $B(y) \sim
(1-y)$, and $C(y) \sim (2-y)\sqrt{1-y}$.
Here $Q^2$ and $y$ are respectively the negative squared 
four-momentum and the fractional energy of the virtual photon, 
$x$ the Bjorken scaling variable and $z$ the fractional energy of the produced hadron.
\\

Among possible mechanisms, two are expected to give important contributions to 
the azimuthal dependence of the unpolarized cross section in the hadron
transverse momentum range accessible at \hermes\.
The first one is called the {\em Cahn effect}~\cite{Cahn:1978se,Cahn:1989yf}, 
a pure kinematic effect where the azimuthal modulations are 
generated by the non-zero intrinsic transverse motion of quarks. 
In the second mechanism, the {\em Boer-Mulders effect}~\cite{Boer:1997nt}, 
$\cos\phi _h$ and $\cos2\phi _h$ modulations originate from the coupling of the quark intrinsic transverse 
momentum and intrinsic transverse spin, a kind of spin-orbit effect.

\section{The \hermes\ experiment}
The results presented here are extracted from data collected at \hermes\ in the $2000-2007$, data taking periods.
The fixed-target \hermes\ experiment ran for more than $10$
years until $2007$ at the electron-positron storage ring of \hera\ at \desy. 
The \hermes\ spectrometer~\cite{Ackerstaff:1998av} was a forward-angle instrument 
consisting of two symmetric (top, bottom) halves above and below the horizontal plane 
defined by the lepton beam pipe. It was characterized by very high 
efficiency (about $98-99\%$) in electron-hadron separation, 
provided by a transition radiation detector, a preshower 
scintillation counter and an electromagnetic calorimeter. 
In addition, a dual-radiator Ring-Imaging CHerenkov (RICH) detector provided 
hadron identification for momenta above 2 GeV/c.

\section{Multi-dimensional unfolding}\label{aba:sec3}
In order to study the new structure functions 
$F_{UU}^{\cos\phi _h}$ and $F_{UU}^{\cos2\phi _h}$ 
defined in Eq.~(\ref{eq:noncol_csec}), 
a measure of the azimuthal modulation of the unpolarized cross section is needed, which
can be extracted via the so-called $\langle{\cos n\phi _h}\rangle$-moments:
\begin{equation}\label{eq:moments}
\langle {\cos n\phi _h}\rangle \,=\,\frac{\int \cos n\phi _h\,d\sigma\,d\phi _h}{\int d\sigma\,d\phi _h}
\end{equation}
with $n=1,2$ and $d\sigma$ defined in equation~\ref{eq:noncol_csec}.

The extraction of these cosine moments from data 
is challenging because they couple to a number of 
{\em experimental sources} of azimuthal modulations, {\em e.g.} detector
geometrical acceptance and higher-order QED effects ({\em radiative effects}). 
Moreover, in the typical case, the event sample is binned 
only in one variable ($1$-dimensional analysis),
and integrated over the full range of all the other ones, but 
the mentioned structure functions and the instrumental spurious contributions
depend on all the kinematic variables $x$, $y$, $z$ and $P_{h\perp}$ simultaneously.
Therefore a $4$-dimensional analysis is needed to take into account
the correlations between the physical modulations and those
spurious contributions, where the event 
sample is binned simultaneously in all the relevant variables~\footnote{For 
a more detailed discussion about one- and multi-dimensional 
analysis see~\cite{Giordano:Transv08}.}.
Therefore, a detailed Monte Carlo simulation of the experimental
apparatus including radiative effects is used to define a $4$-D unfolding
procedure~\cite{Cowan} that corrects the extracted cosine moments
for radiative and instrumenthal effects.

The $4$-D unfolded yields are fit to the functional form:
\begin{equation} \label{eqn:ABC}
{\mathcal A} (1+ {\mathcal B}\csm + {\mathcal C}\cstm)
\end{equation}
where  ${\mathcal B}=2\cs$ and ${\mathcal C}=2\cst$ represent the
desired moments.
One moment pair ($2\cs$, $2\cst$) for each of the $4$-D kinematic
bins is extracted, and the moment dependences on a 
single kinematic variable is obtained projecting the $4$-D results onto the variable under
study by weighting the moment in each bin with the corresponding 
$4\pi$ cross section obtained from a Monte Carlo 
calculation~\footnote{Details on the full $4$-D unfolding and
extraction procedure as well as on the projection versus the single
variable can be found in~\cite{Giordano:2009hi}.}.

\begin{figure}
\centering
\parbox{11cm}{
\includegraphics[width=1\linewidth]{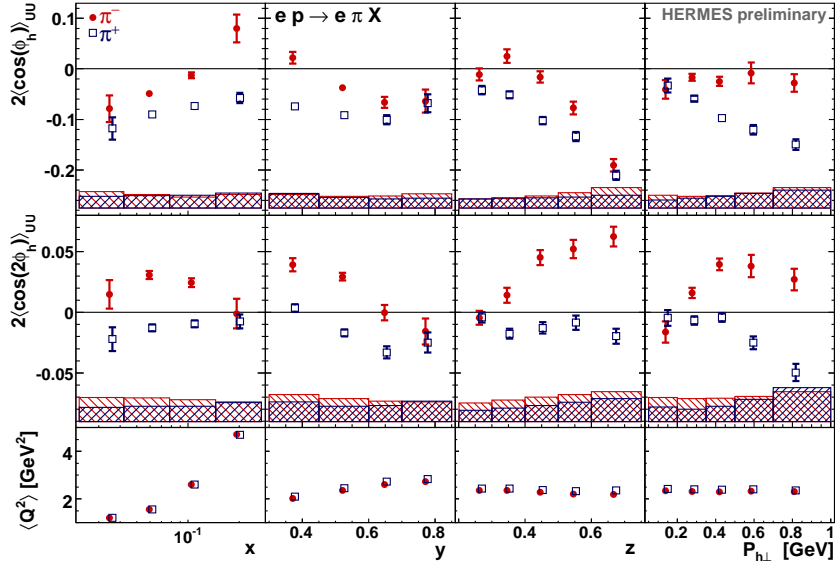}}
\caption{In the upper (middle) panel the $\cs$ ($\cst$) moments for positive (open squares) and negative 
(closed circles) pions, extracted from hydrogen data, projected versus
the kinematic variables $x$, $y$, $z$ and $P_{h\perp}$ are shown. The lower
panel contains the $\langle Q^2 \rangle$ values for each bin}
\label{fig:pions}
\end{figure}
\begin{figure}
\vspace{-1cm}
\centering
\includegraphics[width=0.9\linewidth]{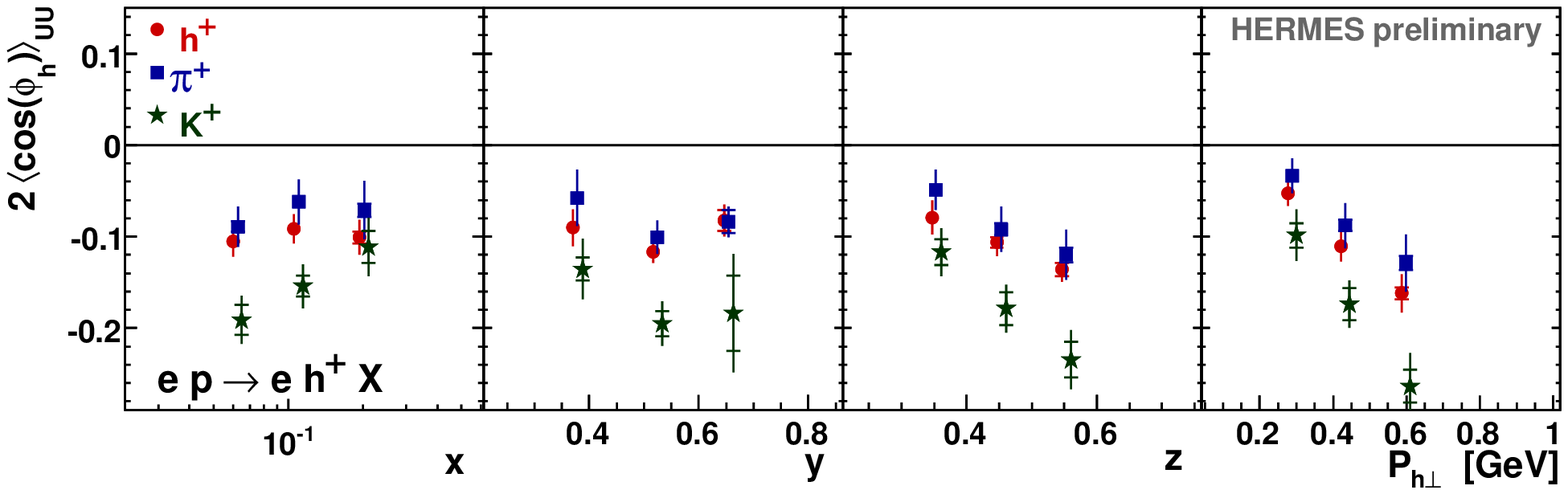}
\includegraphics[width=0.9\linewidth]{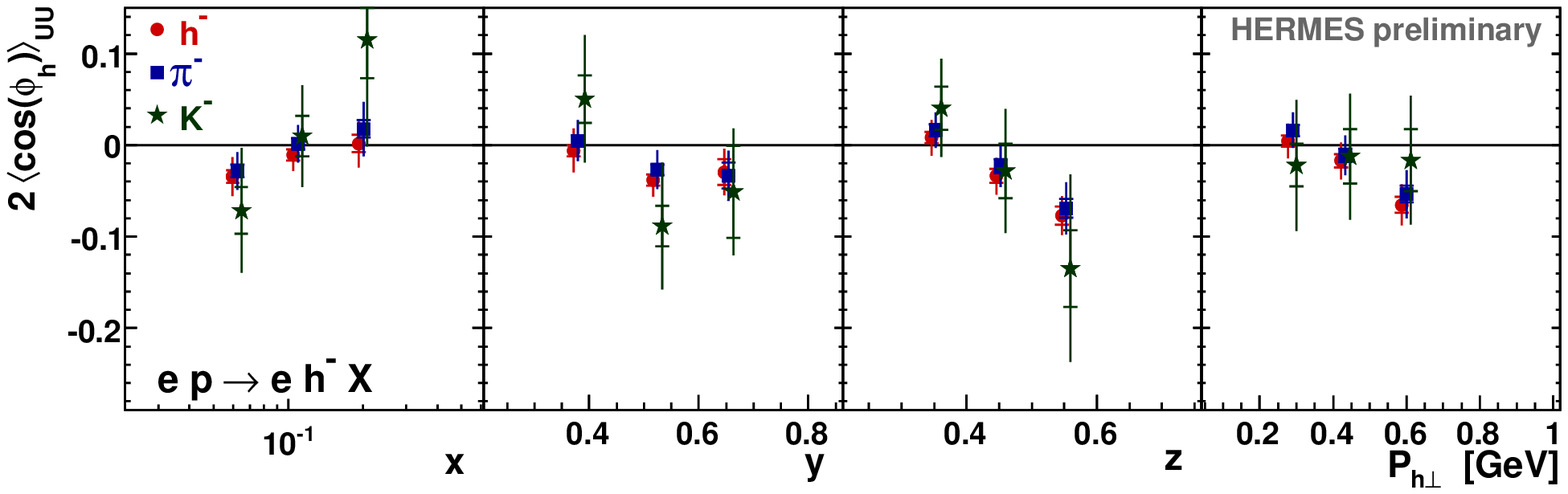}
\caption{$\cs$ moments for positive (upper panel) and negative 
(lower panel) hadrons, extracted from hydrogen data projected versus
the kinematic variables $x$, $y$, $z$ and $P_{h\perp}$.}
\label{fig:cosK}
\end{figure}
\begin{figure}
\centering
\includegraphics[width=0.9\linewidth]{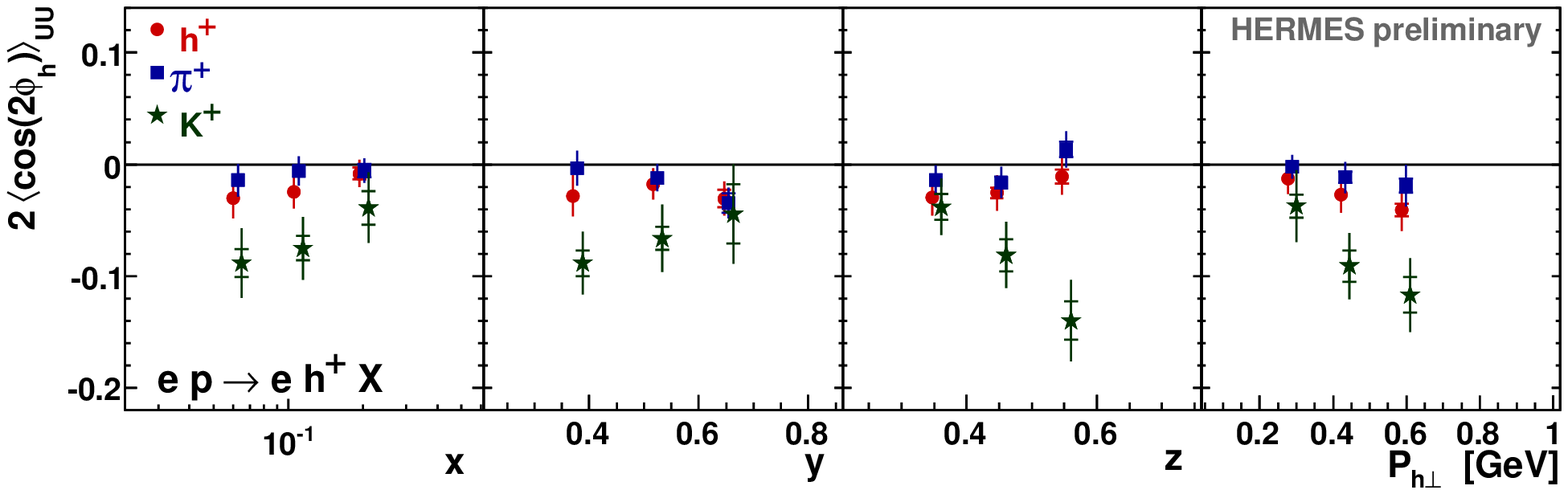}
\includegraphics[width=0.9\linewidth]{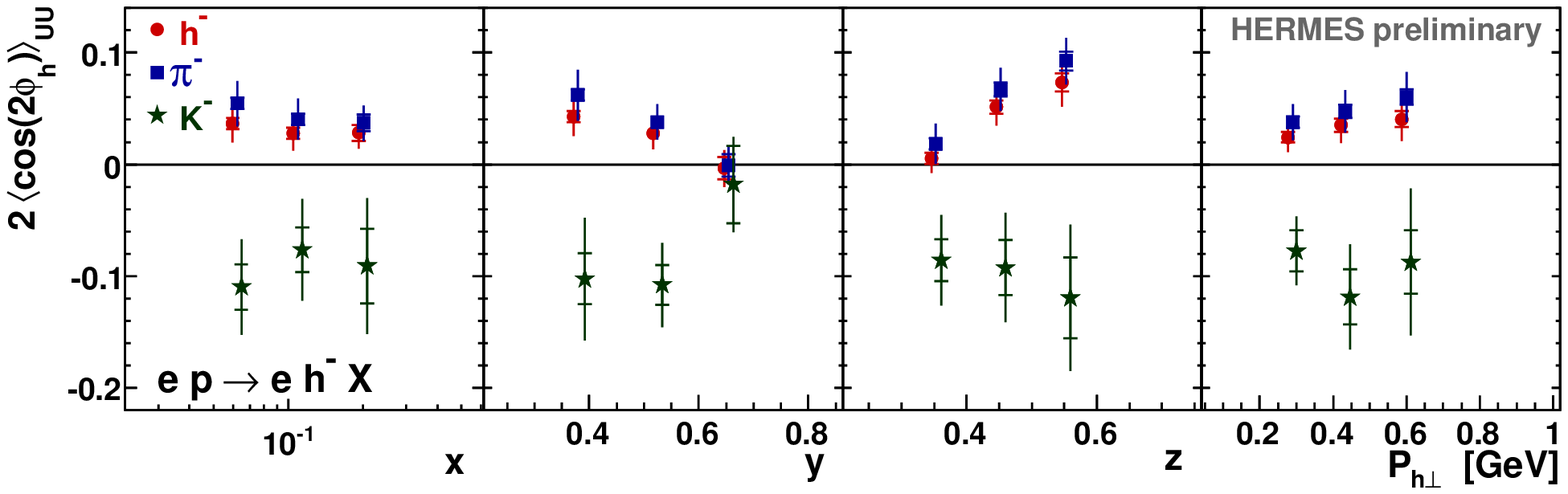}
\caption{$\cst$ moments for positive (upper panel) and negative 
(lower panel) hadrons, extracted from hydrogen data projected versus
the kinematic variables $x$, $y$, $z$ and $P_{h\perp}$.}
\label{fig:cos2K}
\end{figure}

\section{Results}
The cross section unintegrated over hadron transverse momentum 
gives access to new exciting aspects of the nucleon structure, 
which are currently under intense theoretical investigations.

The cross section unintegrated over hadron transverse momentum gives
access to new exciting
aspects of the nucleon structure, which are currently under intense
theoretical investigations.
However, as the extraction of unpolarized cosine moments is
experimentally challenging, very few measurements have been 
performed to date, and, with exception of \compass~\cite{Sbrizzai}, most of them
average out any possible flavor dependence~\cite{Aubert,Arneodo,Breitweg,Adams,Osipenko}.
To date, this analysis at \hermes\ represents
the most complete data set on the subject, and allows access to flavor
dependent information on
the nucleon internal transverse degrees of freedom.

Pions moments projected in the relevant kinematic variables 
are shown in figure~\ref{fig:pions}  for $\cs$ (upper panel) and
$\cst$ (middle panel) moments.
The $\cs$ moments are found to be negative for both charged pions, 
but larger in magnitude for positive ones, while the $\cst$ moments show opposite 
sign for positive and negative pions: both modulations are clearly
charge dependent, and this feature is considered as an evidence of a non-zero Boer-Mulders 
effect~\cite{Gamberg:2007wm,Barone:2008tn,Zhang:2008ez,Barone:2009hw}.

Results for kaons are shown in figure~\ref{fig:cosK}  for $\cs$ moments
and figure~\ref{fig:cos2K}  for $\cst$ moments.
The upper panels show results for positive kaons (stars)
compared to positive pions (squares) and unidentified hadrons (circles) results,
the lower panels the same comparison for negatively charged kaons,
pions and hadrons. Due to the poorer statistics of kaon event samples,
the kaon results have been extracted in a reduced kinematic region
with respect to the pion moments discussed above
resulting in a smaller number of bins shown in the pictures.
The positive kaon $\cs$ moments (figure~\ref{fig:cosK}) are found 
to be negative and larger in magnitude than $\cs$ moments extracted for pions, while
negative kaons behave similarly to negative pions, showing results compatible with zero.

The absolute value of  kaon $\cstm$ modulations are found to be larger
in magnitude than pions ones. While pion $\cstm$ modulations change sign 
between differently charged pions, kaon's modulations are negative for
both kaon charges.
In general the hadrons have a similar trend as the pions but, particularly for the $\cst$ moments, 
the hadrons are shifted to lower values than the pions.
The discrepancy between hadrons and pions is consistent with the observed kaon moments. \\

The cosine modulations have been extracted also for data collected with
deuterium target, and they are found to be compatible with hydrogen
results, both for unidentified hadrons, pions and kaons. 
This suggests that similar contributions arise from $up$ and $down$ 
quarks to the cosine modulations.

\end{document}